\documentclass[conference,10pt,letterpaper]{IEEEtran}
\makeatletter
\def\endthebibliography{%
  \def\@noitemerr{\@latex@warning{Empty `thebibliography' environment}}%
  \endlist
}
\makeatother

\usepackage{cite}

\usepackage{graphicx}

\usepackage{bm}
\usepackage{amsfonts}

\usepackage{amssymb}

\usepackage{color}

\usepackage[outdir=./]{epstopdf}
\ifCLASSINFOpdf
\else
\fi
\usepackage{amsmath}
\usepackage{algorithm}
\usepackage{algorithmic}

\usepackage{booktabs}

\ifCLASSOPTIONcompsoc
 \usepackage[caption=false,font=normalsize,labelfont=sf,textfont=sf]{subfig}
\else
 \usepackage[caption=false,font=footnotesize]{subfig}
\fi
\usepackage{stfloats}
\fnbelowfloat
\newcommand{\CLASSINPUTtoptextmargin}{0.9in}

\begin{document}
%
\title{Dual-functional Radar-Communication Waveform Design under Constant-modulus and Orthogonality Constraints}

\author{\IEEEauthorblockN{Fan Liu, Christos Masouros and Hugh Griffiths
}
\IEEEauthorblockA{Department of Electronic and Electrical Engineering, University College London, London, UK}
Email: $\left\{\text{fan.liu, c.masouros, h.griffiths} \right\}$@ucl.ac.uk
}


%


\maketitle

\begin{abstract}
In this paper, we focus on constant-modulus waveform design for the dual use of radar target detection and cellular transmission. As the MIMO radar typically transmits orthogonal waveforms to search potential targets, we aim at jointly minimizing the downlink multi-user interference and the non-orthogonality of the transmitted waveform. Given the non-convexity in both orthogonal and CM constraints, we decompose the formulated optimization problem as two sub-problems, where we solve one of the sub-problems by singular value decomposition and the other one by the Riemannian conjugate gradient algorithm. We then propose an alternating minimization approach to obtain a near-optimal solution to the original problem by iteratively solve the two sub-problems. Finally, we assess the effectiveness of the proposed approach via numerical simulations.
\end{abstract}


%
\IEEEpeerreviewmaketitle

\section{Introduction}
In the coming generation of wireless communication and radar systems, there will be an ever increasing competition over the scarce spectrum resources \cite{radar_spectrum}. Hence, it is favorable to have both functionalities deployed on a single hardware platform with the shared use of the same frequency band. As an emerging research topic, dual-functional radar-communication (DFRC) not only ensures the efficient usage of the spectrum, but also presents novel system designs that can benefit from the cooperation of sensing and communication \cite{survey}. It is expected that such a technique could contribute towards novel military applications such as the multifunction RF systems\cite{multi_RF}.
\\\indent A critical challenge in DFRC is to design dual-functional waveforms that can detect radar targets and transmit useful information simultaneously. Early contributions on this topic mainly focus on temporal and spectral processing, where a typically used waveform is the chirp signal. In \cite{saddik2007ultra}, the quasi-orthogonality of the up and down chirp waveforms has been exploited to differentiate 0 and 1 in the data sequence. By contrast, the authors of \cite{han2013joint} have proposed a simpler approach based on the time-division (TD) framework, in which the chirp signals are employed for radar target detection, while allowing arbitrary modulation formats to be used for communication. In addition to designing a novel waveform from ground-up, an alternative approach would be to adopt existing communication signals for radar detection, e.g. Orthogonal Frequency Division Multiplexing (OFDM) waveform \cite{sturm2011waveform}.
\\\indent As a step further, recent researches propose to exploit the high degrees-of-freedom (DoFs) of the multi-input-multi-output (MIMO) systems for designing the DFRC waveform, where the spatial-domain processing is further taken into consideration. In \cite{hassanien2016dual}, MIMO radar waveforms have been designed such that the communication bits can be embedded into the sidelobes of the transmit beampattern. A similar approach has been taken in \cite{boudaher2016towards}, where the useful information is transmitted by shuffling the radar waveforms across the transmit antenna array. Note that the above waveform designs rely on the assumption of a line-of-sight (LoS) channel between the MIMO radar and the communication users, which are unable to address the more commonly-seen Non-LoS (NLoS) cases. In view of this, the work in \cite{fan_bf} further assumes an NLoS communication channel in the DFRC scenario, where convex optimization techniques are employed for joint beamforming.
\\\indent While the above methodologies provide basic dual-functional capabilities, little efforts have been done towards the more practical constraints such as constant-modulus (CM) waveforms. In order to fully exploit the transmit power, the RF amplifiers equipped on the radar systems are typically required to operate at the saturation region \cite{QCQP_REF}, which may cause serious distortion to signals with high peak-to-average power ratio (PAPR), such as the OFDM mentioned above. To address this issue, the work in \cite{opt_waveform_fan} proposes a DFRC waveform design by imposing the non-convex CM constraint, which is then solved via a branch-and-bound (BnB) algorithm. Nevertheless, the worst-case complexity of such a method is known to be at the exponential order of the size of the antenna array \cite{Tuy2016Convex}, which may prevent its implementation in realistic systems.
\\\indent In this paper, we propose a low-complexity constant-modulus waveform design for the MIMO DFRC system, which detects targets while communicating with multiple downlink users. As the MIMO radar typically transmits spatially orthogonal waveforms for searching potential targets in the whole angle domain, we firstly minimize the downlink multi-user interference (MUI) under waveform orthogonality constraint, where the optimal waveform can be obtained in closed-form. To attain a flexible trade-off between radar and communication, we further consider a weighted optimization problem by imposing CM constraints, in which the MUI energy and the degrees of non-orthogonality of the waveform are jointly minimized. While the formulated problem is non-convex and generally NP-hard, we propose an alternating minimization (AltMin) algorithm based on manifold optimization techniques. By doing so, a near-optimal solution can be efficiently obtained. Numerical results show that the proposed AltMin approach outperforms the communication-only zero-forcing (ZF) precoding while achieving a favorable performance trade-off between radar and communication.
\begin{figure}[h]
    \centering
    \includegraphics[width=0.6\columnwidth]{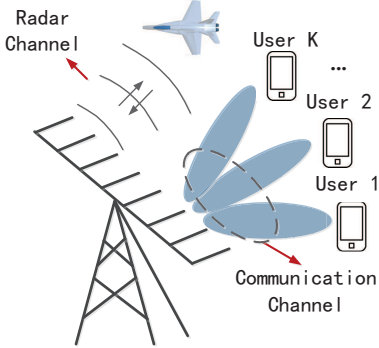}
    \caption{MIMO dual-functional radar-communication system.}
    \label{fig:1}
\end{figure}
\section{System Model}
We consider a MIMO DFRC system equipped with an \emph{N}-antenna uniform linear array (ULA) as shown in Fig. 1, which serves \emph{K} single-antenna users while detecting radar targets in the same time. Below we briefly introduce the system models for both communication and radar functionalities.
\subsection{Communication Model}
The received signal matrix at the downlink users can be obtained as
\begin{equation}\label{eq1}
    {{\mathbf{Y}}} = {\mathbf{H}}{\mathbf{X}} + {\mathbf{Z}},
\end{equation}
where ${\mathbf{X}} = \left[ {{{\mathbf{x}}_1},{{\mathbf{x}}_2},...,{{\mathbf{x}}_L}} \right] \in {\mathbb{C}^{N \times L}}$ represents the transmitted signal matrix, with $L$ being the length of the communication frame/radar pulse, $\mathbf{H} = {\left[ {{{\mathbf{h}}_1},{{\mathbf{h}}_2},...,{{\mathbf{h}}_K}} \right]^T}\in {\mathbb{C}^{K \times N}}$ denotes the channel matrix which is assumed to be perfectly estimated by the DFRC, and ${\mathbf{Z}} = \left[ {{{\mathbf{z}}_1},{{\mathbf{z}}_2},...,{{\mathbf{z}}_L}} \right] \in {\mathbb{C}^{K \times L}}$ is the noise matrix, with ${{\mathbf{z}}_l}\sim\mathcal{C}\mathcal{N}\left( {0,{N_0}{\mathbf{I}}_N} \right),l = 1,2,...,L$.
\\\indent For a given constellation symbol matrix ${\mathbf{S}}\in {\mathbb{C}^{K \times L}}$ that is desired by downlink users, the received signals can be recast in the form
\begin{equation}\label{eq2}
    {{\mathbf{Y}}} = {\mathbf{S}} + \underbrace {\left( {{\mathbf{HX}} - {\mathbf{S}}} \right)}_{{\text{MUI}}} + {\mathbf{Z}}.
\end{equation}
It is assumed that each entry of ${\mathbf{S}}$ is drawn from the same alphabet, e.g. a QPSK constellation. Note that the second term in (\ref{eq2}) represents the MUI signals that interfere the symbol demodulation, with its total energy being measured as
\begin{equation}\label{eq3}
    {P_{{\text{MUI}}}} = \left\| {{\mathbf{HX}} - {\mathbf{S}}} \right\|_F^2.
\end{equation}
Particularly, the communication channel degenerates to a simple additive white Gaussian noise (AWGN) channel if ${P_{{\text{MUI}}}} = 0$. Following \cite{larsson_cep}, the average signal-to-interference-plus-noise ratio (SINR) for downlink users can be maximized by minimizing the MUI energy above. We therefore employ (\ref{eq3}) as the communication performance metric in the rest of the paper.
\subsection{MIMO Radar Model}
In contrast to the conventional phased-array radar that transmits phase-shifted versions of a benchmark signal on each element of the antenna array, MIMO radar transmits individual waveforms on each antenna, which offers the advantage of waveform diversity, allowing more DoFs to be exploited for the system design \cite{MIMO_colocated}. Conventionally, MIMO radar firstly transmits spatially orthogonal waveforms that formulate an omni-directional beampattern, which is typically used for searching the potential targets among the whole angle domain \cite{MIMO_colocated,boudaher2016towards}. After that, a directional beampattern is formulated to point to the directions of interest, and thus obtaining more accurate observations. Without loss of generality, we focus on the omni-directional transmission for the radar, in which case the spatial covariance matrix of the transmitted waveforms is
\begin{equation}\label{eq4}
{{\mathbf{R}}_X} = \frac{1}{L}{\mathbf{X}}{{\mathbf{X}}^H} = \frac{{{P_T}}}{N}{{\mathbf{I}}_N},
\end{equation}
where ${{\mathbf{I}}_N}$ denotes the \emph{N}-dimensional identity matrix, and $P_T$ is the total transmit power.
\\\indent In what follows, we design DFRC waveforms based on the aforementioned communication and radar models.
\section{Problem Formulation}
\subsection{A Simple Closed-form Design}
We firstly consider the following optimization problem
\begin{equation}\label{eq5}
\begin{gathered}
  \mathop {\min }\limits_{\mathbf{X}} \;\left\| {{\mathbf{HX}} - {\mathbf{S}}} \right\|_F^2 
  \;\;s.t.\;\;\frac{1}{L}{\mathbf{X}}{{\mathbf{X}}^H} = \frac{{{P_T}}}{N}{{\mathbf{I}}_N}, \hfill \\
\end{gathered}
\end{equation}
in which we minimize the communication MUI while formulating a spatially orthogonal covariance matrix for radar target detection. Due to the orthogonal constraint involved, problem (\ref{eq5}) is non-convex. Fortunately, it can be classified as an orthogonal Procrustes problem (OPP), which has the following globally optimal solution \cite{viklands2006algorithms}
\begin{equation}\label{eq6}
{\mathbf{X}} = \sqrt {{{L{P_T}} \mathord{\left/
 {\vphantom {{L{P_T}} N}} \right.
 \kern-\nulldelimiterspace} N}} {\mathbf{\tilde U}}{{\mathbf{I}}_{N \times L}}{\mathbf{\tilde V}},
\end{equation}
where ${\mathbf{\tilde U\Sigma \tilde V}} = {{\mathbf{H}}^H}{\mathbf{S}}$ denotes the singular value decomposition (SVD) of ${{\mathbf{H}}^H}{\mathbf{S}}$, with $\mathbf{\tilde U}$ and $\mathbf{\tilde V}$ being the matrices that contain the left and right singular vectors respectively, and ${{\mathbf{I}}_{N \times L}}$ is composed by an \emph{N}-dimensional identity matrix and an $N \times \left(L-N\right)$ zero matrix.
\subsection{Constant-modulus Waveform Design}
It is worth noting that (\ref{eq6}) can only guarantee the orthogonality of the waveforms, which is not able to address the constant-modulus design. Moreover, as the strict equality constraint is imposed in problem (\ref{eq5}), the resultant communication MUI might be still high. To overcome these drawbacks, we formulate the following trade-off optimization problem by considering the CM constraint
\begin{equation}\label{eq7}
\begin{gathered}
  \mathop {\min }\limits_{{\mathbf{X}},{\mathbf{U}}} \;\rho \left\| {{\mathbf{HX}} - {\mathbf{S}}} \right\|_F^2 + \left( {1 - \rho } \right)\left\| {{\mathbf{X}} - {\mathbf{U}}} \right\|_F^2 \hfill \\
  s.t.\;\;{\mathbf{U}}{{\mathbf{U}}^H} = \frac{{L{P_T}}}{N}{{\mathbf{I}}_N},\left| {{x_{n,l}}} \right| = \sqrt {{{{P_T}} \mathord{\left/
 {\vphantom {{{P_T}} N}} \right.
 \kern-\nulldelimiterspace} N}} ,\forall n,\forall l, \hfill \\
\end{gathered}
\end{equation}
where $\rho \in \left[0,1\right]$ is a given weighting factor that determines the weights for radar and communication performances in the DFRC system, ${\mathbf{U}} \in \mathbb{C}^{N \times L}$ is a unitary matrix, and ${{x_{n,l}}}$ denotes the $\left(n,l\right)$-th entry for ${\mathbf{X}}$. Intuitively, the second term in the objective function can be viewed as the degrees of non-orthogonality of $\mathbf{X}$, which can be used to trade-off with the communication MUI energy. Due to the imposed orthogonal and CM constraints, problem (\ref{eq7}) is highly non-convex and NP-hard in general, where the globally optimal solution is not obtainable in polynomial time. To this end, we propose an AltMin algorithm to find a near-optimal solution to the problem.
\section{Proposed Approach}
By taking a closer look at problem (\ref{eq7}), we observe that the optimization variables $\mathbf{X}$ and $\mathbf{U}$ are in fact decoupled with each other, which indicates that problem (\ref{eq7}) can be decomposed as two sub-problems that are much easier to tackle. Accordingly, a sub-optimal solution to (\ref{eq7}) can be attained by iteratively solve the sub-problems, which we discuss in the following.
\subsection{Sub-problem for $\mathbf{U}$}
Let us first solve for $\mathbf{U}$ by fixing $\mathbf{X}$, in which case problem (\ref{eq7}) can be rewritten as
\begin{equation}\label{eq8}
\begin{gathered}
  \mathop {\min }\limits_{\mathbf{U}} \;\left\| {{\mathbf{U}} - {\mathbf{X}}} \right\|_F^2 
  \;\;s.t.\;\;{\mathbf{U}}{{\mathbf{U}}^H} = \frac{{L{P_T}}}{N}{{\mathbf{I}}_N}, \hfill \\
\end{gathered}
\end{equation}
which is again an OPP as in (\ref{eq5}) by letting $\mathbf{H} = {{\mathbf{I}}_N}$. Hence, the optimal solution is given by
\begin{equation}\label{eq9}
{\mathbf{U}} = \sqrt {{{L{P_T}} \mathord{\left/
 {\vphantom {{L{P_T}} N}} \right.
 \kern-\nulldelimiterspace} N}} {\mathbf{\bar U}}{{\mathbf{I}}_{N \times L}}{\mathbf{\bar V}},
\end{equation}
where ${\mathbf{\bar U\bar \Sigma \bar V}} = {\mathbf{X}}$ is the SVD for ${\mathbf{X}}$.
\subsection{Sub-problem for $\mathbf{X}$}
We then fix $\mathbf{U}$ and solve for $\mathbf{X}$. Upon letting ${\mathbf{A}} = {\left[ {\sqrt \rho  {{\mathbf{H}}^T},\sqrt {1 - \rho } {{\mathbf{I}}_N}} \right]^T} \in {\mathbb{C}^{\left( {K + N} \right) \times N}}$ and ${\mathbf{B}} = {\left[ {\sqrt \rho  {{\mathbf{S}}^T},\sqrt {1 - \rho } {{\mathbf{U}}^T}} \right]^T} \in {\mathbb{C}^{\left( {K + N} \right) \times L}}$, the objective function in (\ref{eq7}) can be denoted as $\left\| {{\mathbf{AX}} - {\mathbf{B}}} \right\|_F^2$, in which case the sub-problem for $\mathbf{X}$ can be obtained in a compact form as
\begin{equation}\label{eq10}
\begin{gathered}
  \mathop {\min }\limits_{\mathbf{X}} \;\left\| {{\mathbf{AX}} - {\mathbf{B}}} \right\|_F^2 
  \;\;s.t.\;\;\left| {{x_{n,l}}} \right| = \sqrt {{{{P_T}} \mathord{\left/
 {\vphantom {{{P_T}} N}} \right.
 \kern-\nulldelimiterspace} N}} ,\forall n,\forall l. \hfill \\
\end{gathered}
\end{equation}
Further, let us denote ${\mathbf{\tilde A}} = \sqrt {{{{P_T}} \mathord{\left/
{\vphantom {{{P_T}} N}} \right.
\kern-\nulldelimiterspace} N}} {{\mathbf{I}}_L} \otimes {\mathbf{A}}$, ${\mathbf{\tilde x}} = \operatorname{vec} \left( {\mathbf{X}} \right)$ and ${\mathbf{\tilde b}} = \operatorname{vec} \left( {\mathbf{B}} \right)$, the above problem can be simplified as
\begin{equation}\label{eq11}
\begin{gathered}
  \mathop {\min }\limits_{{{\mathbf{\tilde x}}}} \;{\left\| {{\mathbf{\tilde A\tilde x}} - {\mathbf{\tilde b}}} \right\|^2} 
  \;\;s.t.\;\;\left| {{{\tilde x}_i}} \right| = 1 ,\forall i, \hfill \\
\end{gathered}
\end{equation}
where $\otimes$ is the Kronecker product, and ${{{\tilde x}_i}}, i = 1,2,...,NL$ is the \emph{i}-th entry of ${\mathbf{\tilde x}}$. It can be observed that (\ref{eq11}) is a least-squares (LS) problem defined on the \emph{NL}-dimensional complex circle, which is a Riemannian manifold \cite{petersen2006riemannian}. In what follows, we propose a Riemannian conjugate gradient (RCG) algorithm to obtain a sub-optimal solution \cite{CEP_Fan}.
\\\indent Denoting $\mathcal{M}$ the feasible region of (\ref{eq11}), i.e., the circle manifold. Let ${\mathbf{\tilde x}}$ be an arbitrarily given point on $\mathcal{M}$. A \emph{tangent vector} at ${\mathbf{\tilde x}}$ is defined as the vector that is tangential to any smooth curves on $\mathcal{M}$ through ${\mathbf{\tilde x}}$. The set of all tangent vectors associated with ${\mathbf{\tilde x}}$, denoted as ${T_{{\mathbf{\tilde x}}}}\mathcal{M}$, forms the \emph{tangent space}, which is a Euclidean/linear space. Based on \cite{absil2009optimization}, the tangent space for the complex circle manifold can be given as
\begin{equation}\label{eq13}
{T_{{\mathbf{\tilde x}}}}\mathcal{M} = \left\{ {{\mathbf{w}} \in {\mathbb{C}^{NL \times 1}}\left| {\operatorname{Re} \left( {{\mathbf{w}} \circ {{{\mathbf{\tilde x}}}^*}} \right) = {\mathbf{0}}} \right.} \right\},
\end{equation}
where $\left(\cdot\right)^*$ denotes the conjugate operation, and $\circ$ represents the element-wise multiplication.
\\\indent Before presenting the RCG algorithm, let us compute the gradient of the objective function $f\left( {{\mathbf{\tilde x}}} \right) = {\left\| {{\mathbf{\tilde A\tilde x}} - {\mathbf{\tilde b}}} \right\|^2}$, which can be simply given by
\begin{equation}\label{eq14}
\nabla f\left( {{\mathbf{\tilde x}}} \right) = 2{{{\mathbf{\tilde A}}}^H}\left( {{\mathbf{\tilde A\tilde x}} - {\mathbf{\tilde b}}} \right).
\end{equation}
In the RCG framework, (\ref{eq14}) is called the \emph{Euclidean gradient}, which, however, is not the steepest ascent direction on the manifold. Instead, the \emph{Riemannian gradient} is adopted in the iteration of the algorithm, which is defined as the orthogonal projection of (\ref{eq14}) onto the tangent space (\ref{eq13}), and can be given as \cite{absil2009optimization}
\begin{equation}\label{eq15}
  \operatorname{grad} f\left( {{\mathbf{\tilde x}}} \right) = {\mathcal{P}_{{\mathbf{\tilde x}}}}\nabla f\left( {{\mathbf{\tilde x}}} \right)
   \triangleq \nabla f\left( {{\mathbf{\tilde x}}} \right) - \operatorname{Re} \left( {\nabla f\left( {{\mathbf{\tilde x}}} \right) \circ {{{\mathbf{\tilde x}}}^*}} \right) \circ {\mathbf{\tilde x}}, 
\end{equation}
where ${\mathcal{P}_{{\mathbf{\tilde x}}}}\left(\cdot\right)$ is the projector.
\\\indent Note that by stepping towards the negative direction of the Riemannian gradient (\ref{eq15}), the resultant point will be on the tangent space instead of the manifold itself. To obtain the associated point on the circle manifold, a \emph{retraction} mapping is further defined to map a point from ${T_{{\mathbf{\tilde x}}}}\mathcal{M}$ to $\mathcal{M}$. This can be given as the following \cite{absil2009optimization}
\begin{equation}\label{eq16}
{\mathcal{R}_{{\mathbf{\tilde x}}}}\left( {\mathbf{w}} \right) = {\left[ {\frac{{{{\tilde x}_1} + {w_1}}}{{\left| {{{\tilde x}_1} + {w_1}} \right|}},...,\frac{{{{\tilde x}_{NL}} + {w_{NL}}}}{{\left| {{{\tilde x}_{NL}} + {w_{NL}}} \right|}}} \right]^T},
\end{equation}
where $\mathbf{w} \in {T_{{\mathbf{\tilde x}}}}\mathcal{M}$.
\\\indent In the conventional conjugate gradient (CG) algorithm, the descent direction ${\mathbf{d}}_k$ at the \emph{k}-th iteration is the linear combination of the current gradient $\nabla f\left( {{\mathbf{\tilde x}}}_k \right)$ and the previous descent direction ${\mathbf{d}}_{k-1}$, where we have \cite{absil2009optimization}
\begin{equation}\label{eq17}
{{\mathbf{d}}_k} =  - \nabla f\left( {{{{\mathbf{\tilde x}}}_k}} \right) + {\alpha _k}{{\mathbf{d}}_{k - 1}},
\end{equation}
where $\alpha _k$ is a combination coefficient. Nevertheless, such combination is not possible in the RCG algorithm, as the Riemannian gradient $\operatorname{grad} f\left( {{\mathbf{\tilde x}}}_k \right)$ and the descent direction ${\mathbf{d}}_{k-1}$ belong to different tangent spaces, i.e. ${T_{{{{\mathbf{\tilde x}}}_k}}}\mathcal{M}$ and ${T_{{{{\mathbf{\tilde x}}}_{k - 1}}}}\mathcal{M}$, respectively. To resolve this issue, the following non-linear combination is used to compute the descent direction of the \emph{k}-th iteration \cite{absil2009optimization}
\begin{equation}\label{eq18}
{{\mathbf{d}}_k} =  - \operatorname{grad} f\left( {{{{\mathbf{\tilde x}}}_k}} \right) + {\beta _k}{\mathcal{P}_{{{{\mathbf{\tilde x}}}_k}}}{{\mathbf{d}}_{k - 1}},
\end{equation}
where ${\mathbf{d}}_{k-1}$ is projected to ${T_{{{{\mathbf{\tilde x}}}_k}}}\mathcal{M}$, such that its projection can be linearly combined with $\operatorname{grad} f\left( {{\mathbf{\tilde x}}}_k \right)$ located in the same tangent space. The combination coefficient ${\beta _k}$ is computed following the Polak-Ribi\'ere rule \cite{absil2009optimization}.
\\\indent Based on above, we summarize the RCG algorithm for solving (\ref{eq11}) in Algorithm 1.
\\\indent {\emph{Remark:}} As the strict convergence analysis of RCG still remains open problem, it is rather intractable to derive the maximum iteration number needed given an accuracy threshold $\varepsilon$ \cite{absil2009optimization}. We therefore show the complexity per iteration of the algorithm. By simple calculation, it can be proven that each iteration of Algorithm 1 involves $\mathcal{O}\left(NKL\right)$ complex multiplications. The total complexity is therefore $\mathcal{O}\left(N_{iter}NKL\right)$, where $N_{iter}$ is the number of iterations.
\renewcommand{\algorithmicrequire}{\textbf{Input:}}
\renewcommand{\algorithmicensure}{\textbf{Output:}}
\begin{algorithm}
\caption{RCG Algorithm for Solving (\ref{eq11})}
\label{alg:C}
\begin{algorithmic}
    \REQUIRE $\mathbf{H},\mathbf{S},{\mathbf{U}}$, weighting factor $0 \le \rho \le 1$, $P_T$, tolerable accuracy $\varepsilon > 0$, maximum iteration number $k_{max} > 2$
    \ENSURE $\mathbf{\tilde x}_{k}$
    \STATE 1. Compute $\mathbf A$, $\mathbf B$. Initialize randomly $\mathbf{\tilde x}_{0} = \mathbf{\tilde x}_{1} \in \mathcal{M}$, set $\mathbf{d}_0=-\operatorname{grad}f\left(\mathbf{\tilde x}_{0}\right)$, $k=1$.
    \WHILE{$k\le k_{max}$ and ${\left\| {\operatorname{grad} f\left( {{{{\mathbf{\tilde x}}}_{k}}} \right)} \right\|} \ge {\varepsilon}$}
    \STATE 2. Compute the combination coefficient $\beta_k$ using the Polak-Ribi\'ere formula \cite{absil2009optimization}.
    \STATE 3. Compute the descent direction ${\mathbf{d}_k}$ by (\ref{eq18}).
    \STATE 4. Compute stepsize $\mu_{k}$ by the Armijo line search method, and set ${{{\mathbf{\tilde x}}}_{k+1}}$ by
    \begin{equation*}
    {{{\mathbf{\tilde x}}}_{k+1}} = {\mathcal{R}_{{{\mathbf{\tilde x}}_{k}}}}\left( {{\mu_k}{{\mathbf{d }}_k}} \right).
    \end{equation*}
    \STATE 5. $k = k + 1.$
    \ENDWHILE
\end{algorithmic}
\end{algorithm}
\subsection{The Alternating Minimization Procedure}
We are now ready to describe the proposed AltMin method. As shown in Algorithm 2, we repeatedly solve the two aforementioned sub-problems. The algorithm terminates once a preset tolerable accuracy threshold is reached. It is worth highlighting that Algorithm 2 can be viewed as a coordinate descent method, and hence its convergence can be strictly guaranteed \cite{altmin}. In our simulations, we see that Algorithm 2 always converges within tens of iterations within a modest accuracy.
\renewcommand{\algorithmicrequire}{\textbf{Input:}}
\renewcommand{\algorithmicensure}{\textbf{Output:}}
\begin{algorithm}
\caption{Alternating Minimization Algorithm for (\ref{eq7})}
\label{alg:A}
\begin{algorithmic}
    \REQUIRE $\mathbf{H},\mathbf{S}$, $0 \le \rho \le 1$, $P_T$, tolerable accuracy $\eta > 0$ and the maximum iteration number $n_{\max}$
    \ENSURE ${\mathbf{X}}_n$, $\mathbf{U}_n$
    \STATE 1. Initialize randomly ${\mathbf{X}}_{0}$ and ${\mathbf{U}}_{0}$. Compute the objective function of (\ref{eq7}), denoted as $f_{0}$. Set $n=1$.
    \WHILE{$n\le n_{max}$ and $ \left| {{f_{n}} - {f_{n - 1}}} \right|\ge {\eta }$}
    \STATE 2. Compute ${\mathbf{U}}_{n}$ by (\ref{eq9}).
    \STATE 3. Compute ${\mathbf{X}}_{n}$ using Algorithm 1.
    \STATE 4. Compute the objective function ${f_{n}}$ based on the obtained variables.
    \STATE 5. $n = n + 1.$
    \ENDWHILE
\end{algorithmic}
\end{algorithm}

\section{Numerical Results}
\begin{figure}
    \centering
    \includegraphics[width=\columnwidth]{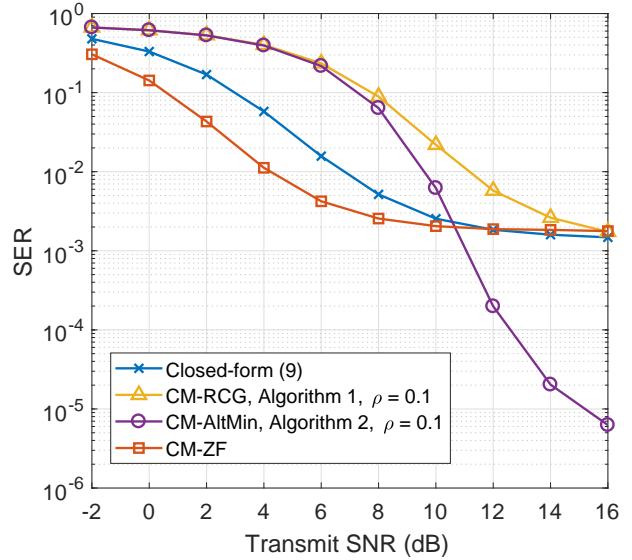}
    \caption{Symbol error rate for downlink users, $N = 16, K = 4$.}
    \label{fig:2}
\end{figure}
In this section, we validate the effectiveness of the proposed approach by numerical simulations, where a MIMO DFRC system equipped with $N = 16$ antennas serves $K = 4$ users in the downlink. Without loss of generality, the spacing between adjacent antennas is set as half-wavelength. The transmitted symbols in $\mathbf{S}$ are drawn from a normalized QPSK constellation. For convenience, we set the total transmit power as $P_T = 1$, and assume a Rayleigh flat-fading channel for the downlink, where each entry of $\mathbf{H}$ subjects to standard complex Gaussian distribution.
\\\indent We first show in Fig. 2 the communication performance in terms of the symbol error rate (SER), where the results of different methods are present for comparison. In particular, we use `Closed-form' for the orthogonal waveform design in (\ref{eq9}), `CM-RCG' for the RCG method with an arbitrarily given unitary matrix, `CM-AltMin' for the proposed AltMin approach by iteratively obtaining the waveform matrix $\mathbf{X}$ and the unitary matrix $\mathbf{U}$, and finally 'CM-ZF' for the conventional ZF precoding design. For the sake of fairness, the modulus of the ZF-precoded waveform is normalized to be constant. While the proposed AltMin method ensures both CM constraints and the quasi-orthogonality of the waveform, we see that it still considerably outperforms both the CM-ZF and the closed-form design (\ref{eq9}) when the target SER is below $10^{-3}$, even with a very small weight $\rho = 0.1$ at the communication's side. It is also reasonable to see that CM-RCG performs worse than the closed-form and the AltMin designs, as it does not update the unitary matrix $\mathbf{U}$, which leads to a high MUI energy.
\\\indent In Fig. 3, we further investigate the detection performance of the proposed waveform designs by using detection probability $P_D$ as a metric, where the classic constant false-alarm rate (CFAR) detection is employed with a fixed false-alarm rate $P_{FA} = 10^{-7}$. In the considered scenario, a point-like target is placed at the far-field, with an azimuth angle of $\theta = 20^\circ$. The detection probability is computed following [19, eq. (69)]. It is not surprising to observe that the closed-form design (9) yields the best detection performance, as the orthogonal signal is known to be the optimal searching waveform of the MIMO radar \cite{NSP_MIMO}. By further looking at the performance of the proposed AltMin method, we see that it shows significant gain over the CM-ZF precoding design. Moreover, the performance-loss of the AltMin approach comparing with the closed-form design is less than 1dB, which again proves its superiority relative to other techniques. Together with Fig. 2, it is noteworthy that the AltMin algorithm can achieve a favorable performance trade-off between radar and communication while guaranteeing the CM constraints for the transmitted waveform.
\begin{figure}
    \centering
    \includegraphics[width=\columnwidth]{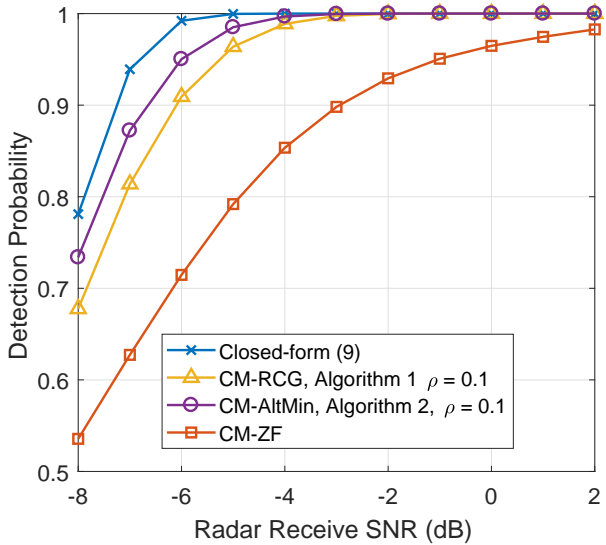}
    \caption{Detection probability under different waveform designs, $N = 16, K = 4, P_{FA} = 10^{-7}$.}
    \label{fig:3}
\end{figure}

\section{Conclusion}
This paper has investigated the constant-modulus (CM) waveform design for the dual-functional radar-communication (DFRC) system. To minimize the communication multi-user interference (MUI) while preserving the orthogonality of the waveform for radar detection, we have proposed a simple closed-form design by singular value decomposition (SVD). To further improve the performance and to avoid signal distortion in the non-linear power amplifier, we have jointly minimized the weighted summation of the MUI and the non-orthogonality for the waveform under non-convex CM constraints. The proposed optimization problem can be solved via Riemannian conjugate gradient (RCG) and the alternating minimization (AltMin) approaches. Simulation results have been presented to compare the performance of the proposed method and several benchmark techniques, which indicate that the AltMin algorithm attains a favorable performance trade-off between radar and communication functionalities while guaranteeing the CM constraints.





%
\bibliographystyle{IEEEtran}
\bibliography{IEEEabrv,SSPD}

\end{document}